# Gate controlled large resistance switching driven by charge density wave in 1T-TaS$_2$/2H-MoS$_2$ heterojunction

Mehak Mahajan[∥], Krishna Murali[∥], Nikhil Kawatra, and Kausik Majumdar[*]

Department of Electrical Communication Engineering, Indian Institute of Science, Bangalore 560012, India

[∥]These authors contributed equally, [*]*Corresponding author,* email: kausikm@iisc.ac.in

**ABSTRACT:** 1T-TaS$_2$ is a layered material that exhibits charge density wave (CDW) induced distinct electrical resistivity phases and has attracted a lot of attention for interesting device applications. However, such resistivity switching effects are often weak, and cannot be modulated by an external gate voltage – limiting their widespread usage. Using a back-gated 1T-TaS$_2$/2H-MoS$_2$ heterojunction, here we show that the usual resistivity switching in TaS$_2$ due to different phase transitions is accompanied with a surprisingly strong modulation in the Schottky barrier height (SBH) at the TaS$_2$/MoS$_2$ interface – providing an additional knob to control the degree of the phase-transition-driven resistivity switching by an external gate voltage. In particular, the commensurate (C) to triclinic (T) phase transition results in an increase in the SBH owing to a collapse of the Mott gap in TaS$_2$. The change in SBH allows us to estimate an electrical Mott gap opening of ∼71 ± 7 meV in the C phase of TaS$_2$. On the other hand, the nearly-commensurate (NC) to incommensurate (IC) phase transition results in a suppression in the SBH, and the heterojunction shows a gate-controlled resistivity switching up to 17.3, which is ∼14.5 times higher than that of standalone TaS$_2$. The findings mark an important step forward showing a promising pathway to externally control as well as amplify the CDW induced resistivity switching. This will boost device applications that exploit these phase transitions, such as ultra-broadband photodetection, negative differential conductance, fast oscillator and threshold switching in neuromorphic circuits.

Transition metal dichalcogenides are emerging out as pertinent materials for applications in high performance flexible electronics and optoelectronics[1,2,3]. 1T-TaS$_2$ is a distinct layered material that exhibits multiple conductivity phases resulting from strong electron-phonon and electron-electron interactions. This material hosts a wide variety of charge density wave (CDW)[4,5] states and the CDW amplitude is significant in 1T-polytype as compared to 2H-polytype of TaS$_2$[6,7]. Depending on the temperature of the 1T-TaS$_2$ crystal, CDW exists in different phases subject to the alignment with the underlying lattice[6,8,9,10]. As the temperature is reduced below 550 K, the metallic crystal undergoes a CDW phase transition, however the CDW remains incommensurate (IC) with the underlying crystal lattice. On further cooling, it undergoes IC to nearly-commensurate (NC) CDW phase transition at 340 K, and an NC to commensurate (C) phase transition at 180 K. The NC to C transition is accompanied with a Mott transition resulting in strong suppression of conductivity. On heating, the crystal undergoes a new phase transition, from C phase to the triclinic (T) phase appearing at 223 K[11,12,13], followed by a T to NC, and NC to IC phase transitions at 283 K and 353 K, respectively. The different CDW phase transitions can also be controlled by pressure[8,14], doping[15,16], thickness[17,18] and photoexcitation [9,19].

Among these different phase transitions, the C-T transition is of great scientific interest due to a large resistivity switching of more than an order of magnitude[7,8,18] owing to a Mott gap opening associated with the phase transition. While the Mott gap opening has been extracted by several reports using various optical techniques[9,10,20,21,22,23,24,25], an estimation of the electrical gap directly from transport measurement of a TaS$_2$ device is missing. On the contrary, the NC-IC transition can be electrically driven while operating at room temperature and thus has attracted a lot of attention in device applications, including wideband photodetector[26], fast oscillator[27] and neuromorphic circuits[28]. However, there are two intrinsic bottlenecks with such resistivity switching. First, the resistivity switching ratio during the NC-IC phase transition is quite weak (< 2)[7,8,18]. Second, the phase transition driven resistivity switching of $TaS_2$ cannot be controlled by an external gate voltage. Hence, improving the switching ratio in the NC-IC phase transition and adding a possible gate controllability would be of great importance for the advancement of these applications.

In order to address these issues, in this work, we employ a 1T-$TaS_2$/2H-$MoS_2$ heterojunction (lattice mismatch of ~6.33%[29,30]) in a back gated field effect transistor (FET) structure. We show that the $TaS_2$/$MoS_2$ interface exhibits a low barrier, high performance van der Waals (vdW) electrical contact[31,32,33,34,35,36], which is promising for pathway towards "all-2D" flexible devices. We also demonstrate that both the C-T and NC-IC phase transitions not only result in a change in the resistivity in the $TaS_2$ film, but also bring about a change in the Schottky barrier height (SBH) at the $TaS_2$/$MoS_2$ interface. This allows us to control the phase transition driven carrier transport through the heterojunction device by the application of a gate voltage. The C-T phase transition results in an increase in the SBH, which allows us to electrically estimate the Mott gap opening in 1T-$TaS_2$ at the C phase. On the other hand, the NC-IC phase transition reduces the SBH at the $TaS_2$/$MoS_2$ interface, which, depending on the gate voltage applied, enhances the switching ratio by a factor as much as 14.5X compared with 1T-$TaS_2$ control.

**Results and Discussions:**

2H-$MoS_2$ is a layered transition metal di-chalcogenide (TMDC) semiconductor, which is appealing as a channel material in electronic device applications owing to its appreciable bandgap, moderate carrier mobility, and channel length scalability. The heterojunction device used in this work is schematically shown in Fig. 1a, where a back gated $MoS_2$ channel is formed with asymmetric contacts, namely Ni and $TaS_2$ contacts on two different sides. We note that Ni makes good electrical contact with $MoS_2$[37,38,39] (see **Supplemental Material S1**[40]) owing to efficient interfacial charge transfer resulting from strong hybridization of partially filled Ni-*3d* and S-*3p* orbitals[39]. By switching the polarity of the drain voltage, we study characteristics of the carrier injection through the $TaS_2/MoS_2$ contact interface while taking $Ni/MoS_2$ junction of the same device as the reference interface. To fabricate the heterojunction device, we first exfoliate few-layer $2H-MoS_2$ flakes on 285 nm thick $SiO_2$ coated heavily-doped Si substrate. We next transfer thin layers of $1T-TaS_2$ on top of the $MoS_2$ flake under a microscope using a micromanipulator. The contact electrodes are patterned by electron beam lithography, followed by electron-beam evaporation of Ni (10nm)/Au (50 nm), and subsequent lift-off. Fig. 1b shows the optical image of the device after completion of fabrication. The top panel of Fig. 1c depicts the corresponding thickness mapping image using Atomic Force Microscope (AFM). The bottom panel of Fig. 1c shows the thickness of the $MoS_2$ and the $TaS_2$ flakes are 6.4 nm and 43.6 nm, respectively, as measured along the green dashed arrow. The devices reported in this work are measured multiple times over a period of several weeks and no noticeable degradation of the device characteristics is observed due to surface oxidation and other ambience induced effects.

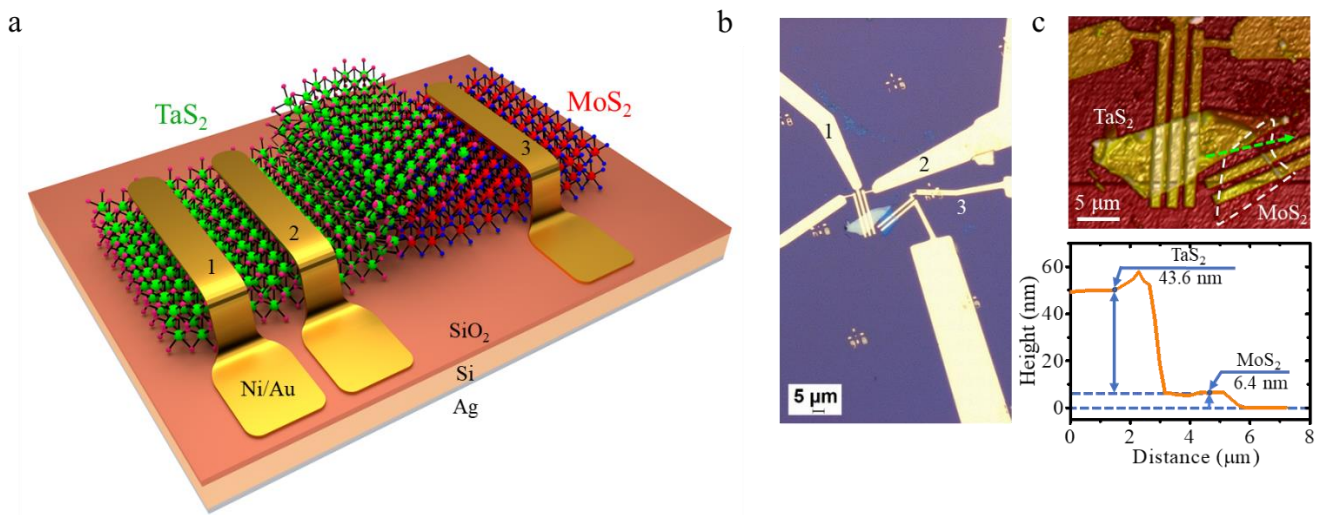

**Figure 1. 1T-TaS$_2$/2H-MoS$_2$ heterojunction.** (a) Schematic of the device with terminals 1 and 2 probing the TaS$_2$ control device (TS12) and terminals 2 and 3 probing the heterojunction device (H23). (b) Optical image of the fabricated device. Scale bar: 5 µm. (c) Top panel: AFM thickness mapping image of the heterojunction device. Scale bar: 5 µm. Bottom panel: Thickness of the MoS$_2$ and TaS$_2$ flakes along the green dashed arrow in the top panel.

Raman spectroscopy is a useful tool to characterize CDW phase transitions in 1T-TaS$_2$[41,42,43,44,45]. When the crystal has not undergone any CDW phase transition, due to the high symmetry of the crystal, specific zone center phonons participate in the first order Raman scattering in order to maintain both energy and momentum conservation. However, once a CDW phase change sets in, the lattice distorts, reducing the translational symmetry of the crystal. This relaxes the condition of first order Raman scattering at the zone center and results in a large number of Raman active vibrational modes[41]. Fig. 2a shows the acquired Raman spectra from 1T-TaS$_2$ in the heating cycle using a 532 nm laser excitation at 193 K (C phase) and 300 K (NC phase), which are in agreement with previous reports[41,42,45]. In the C phase, the distinct Raman peaks at the lower frequencies (between 90 cm$^{-1}$ and 140 cm$^{-1}$) result from acoustic branches and directly correlate with the signature of the commensurate nature of the C phase. The higher frequency peaks (between 200 cm$^{-1}$ to 400 cm$^{-1}$) originate from optical phonons[42], and can be observed in both C and NC phases.

Fig. 2b depicts the resistance (*R*) - temperature (*T*) characteristics of a representative two-probe TaS$_2$ device (in the inset) in vacuum in the heating cycle under small electric field condition. Each layer of 1T-TaS$_2$ crystal structure is composed of tantalum (Ta) atoms, which are surrounded by sulphur atoms in an octahedral arrangement[13]. The high resistance state at low temperature results from the C phase, where the *David-star* structure[6,46], as depicted in Fig. 2c, forms a commensurate structure with the underlying lattice. This commensurate phase results from the inward displacement of the twelve Ta atoms, located at the star corners, towards the thirteenth Ta atom at the center of the star. The atomic displacement results in the deformation in the structure, including a swelling at the star center[30,47]. The reduction in the interatomic distance strengthens the bonds inside the David-star in comparison to the bonds outside the star, resulting in the disintegration of the band structure into submanifolds. The twelve corner Ta atoms of the star contribute electrons to the two three-band submanifolds in the valence band, whereas the thirteenth atom at the center of the star contributes one electron to the submanifold in the conduction band (Fig. 2d – left panel). It has been suggested that the spin-orbit coupling forces further reconstruction in the band structure and results in a unique narrow band at the Fermi level that is partially filled[48,49]. This facilitates electron-electron interaction induced Mott transition in the lattice and a Mott gap opens up (Fig. 2d – right panel)[47,49].

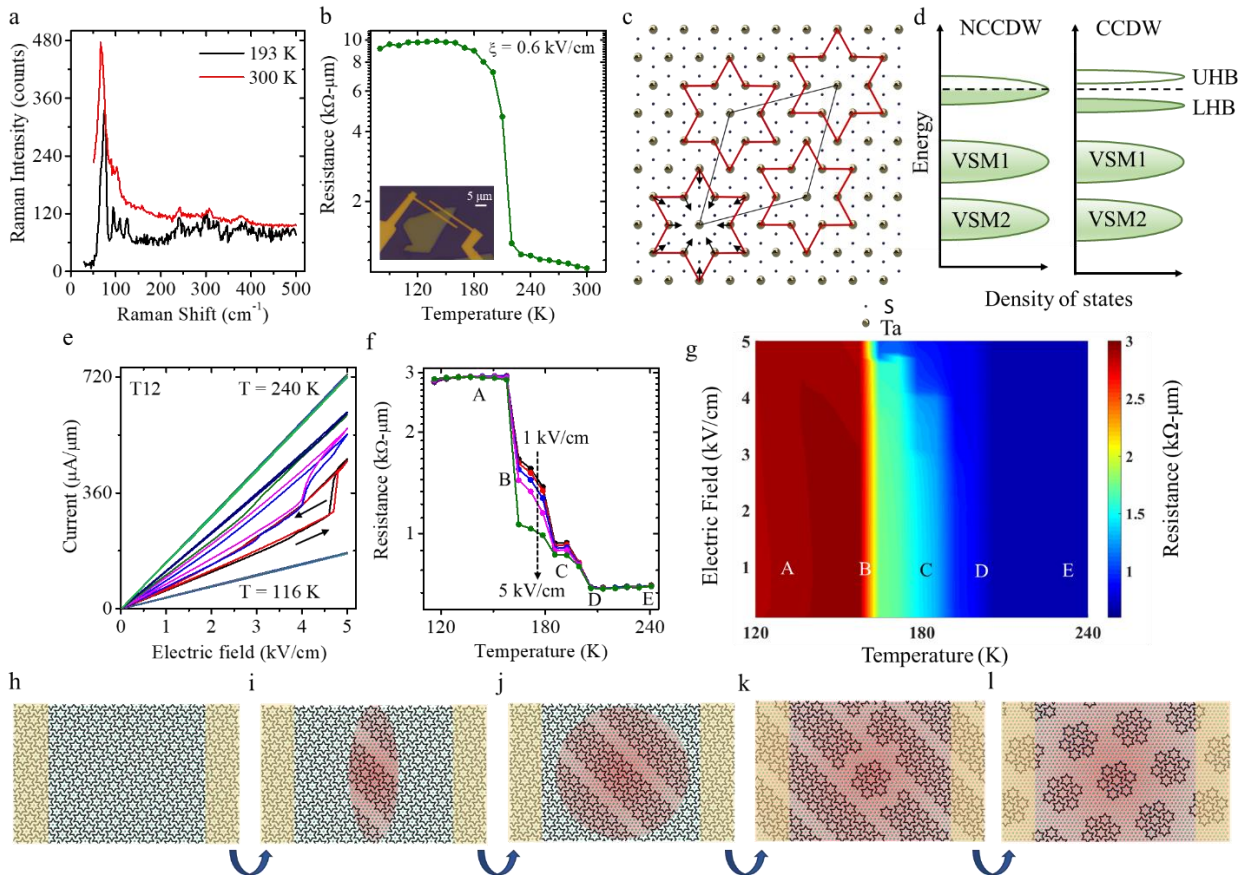

**Figure 2. Electrical tuning of phase transitions in 1T-TaS$_2$.** (a) Raman shift of 1T-TaS$_2$ in NC phase (at 300 K in red) and in C phase (at 193 K in black) during the heating cycle. (b) Temperature dependent resistance of a representative TaS$_2$ flake under low field condition. Inset: Optical image of the device. (c) David star structure formation in the C phase of 1T-TaS$_2$. (d) Hubbard model of 1T-TaS$_2$ depicting Mott gap opening. (e) Current – electric field characteristics of 1T-TaS$_2$ two probe device (TS12 – probing terminals 1 and 2 in Fig. 1a-b) in the temperature range 116 K to 240 K. The black arrows indicate the bias sweep direction. (f) Resistance – temperature plot of the forward sweep in (e) indicating multiple resistance states. (g) Colour plot of resistance in the temperature-field space. (h)-(l) Joule heating induced localized phases of the device giving rise to multiple resistance states in (f).

*Electrically accessing different TaS$_2$ phases:* In order to access the different CDW phases of TaS$_2$ electrically, we next apply a high field across the probes 1 and 2 of the TaS$_2$ device TS12 in Fig. 1a. The hysteretic bi-stable switching, as observed in Fig. 2e, is indicative of external bias controlled phase change of the TaS$_2$ flake. The sharp change in resistance is observed around 160 K (Fig. 2f), which is lower than the C-T phase transition temperature (~220 K) under low field in Fig. 2b. This suggests that Joule heating induced increase in local temperature plays a key role in the phase change. A color phase plot of the different resistance-states in the temperature-electric field space is shown in Fig. 2g.

We construct a simple model for the multi-state resistance switching in the TaS$_2$ flake, as explained in Fig. 2h-l. Initially, at low temperature, the whole flake is in C phase, with linear current-field characteristics (Fig. 2h). This situation is denoted by point A in Fig. 2f-g. As the sample is heated close to the C-T phase transition temperature ($T_{cT}$), the current induced Joule heating drives the local temperature at the central part of the flake (which is farthest from contact heat sinks) at a higher value than the rest of the flake. Note that, in the C phase, particularly, close to $T_{cT}$, an increase in the temperature results in a steep reduction in the lattice component of the thermal conductivity, suppressing the overall thermal conductivity[50]. This provides a positive feedback and further helps to increase the local temperature. Eventually, the temperature of the central part is driven beyond $T_{cT}$, forcing a local C-T phase transition (point B), as schematically depicted in Fig. 2i. This corresponds to a steep jump in the overall resistance of the sample. With further increase in the drain field or heating of the sample (point C), the local temperature of the surrounding portion also increases causing a gradual increase in the size of the central T phase region (Fig. 2j), and, in turn, results in the gradual reduction in the resistance. When the temperature and field are increased further, the whole flake is eventually converted into T phase (Fig. 2k), and no further change in resistance is observed beyond this point (point D). At higher temperature (point E - beyond 283 K lattice temperature), the whole flake transforms into NC phase (Fig. 2l), however, T-NC phase transition has an almost negligible impact on further change in resistance[11,13]. The above-mentioned Joule heating induced phase transition mechanism is qualitatively supported by the hysteresis observed in the current-field plot in Fig. 2e. The flake undergoes a C-T phase transition due to Joule heating in the forward sweep. When the field is withdrawn, Joule heating is suppressed, but the flake does not immediately come back to high resistance C phase until the flake cools down below 180 K, resulting in hysteresis.

***Efficient carrier injection by 1T-TaS₂ contact:*** We next explore the carrier injection efficiency from 1T-TaS$_2$ to 2H-MoS$_2$ in the heterojunction device (H23) shown in Fig. 1a, by probing the terminals 2 and 3. We take TaS$_2$ and Ni as the source (S) and the drain (D), respectively, in the rest of the paper. Thus, owing to the asymmetric design of the device, for $V_d (= V_{32}) > 0$, electrons are injected from the TaS$_2$ contact, while for $V_d < 0$, electrons are injected from Ni into the MoS$_2$ channel. Hence, by switching the polarity of $V_d$, we can probe the carrier injection from individual contacts, as schematically depicted in the insets of Fig. 3a-b. The transfer characteristics of the device for the two cases, shown in Fig. 3a-b at $T = 240$ K (T-phase), indicate an on-off ratio in excess of $10^6$, irrespective of the carriers being injected from Ni or TaS$_2$. Fig. 3c shows the output characteristics at different back gate voltages ($V_g$). We clearly observe that the magnitude of the drive current is similar for both TaS$_2$ and Ni injection cases, suggesting excellent carrier injection efficiency of TaS$_2$/MoS$_2$ interface. Such highly efficient carrier injection from the TaS$_2$/MoS$_2$ junction is promising for vdW-vdW contact engineering.

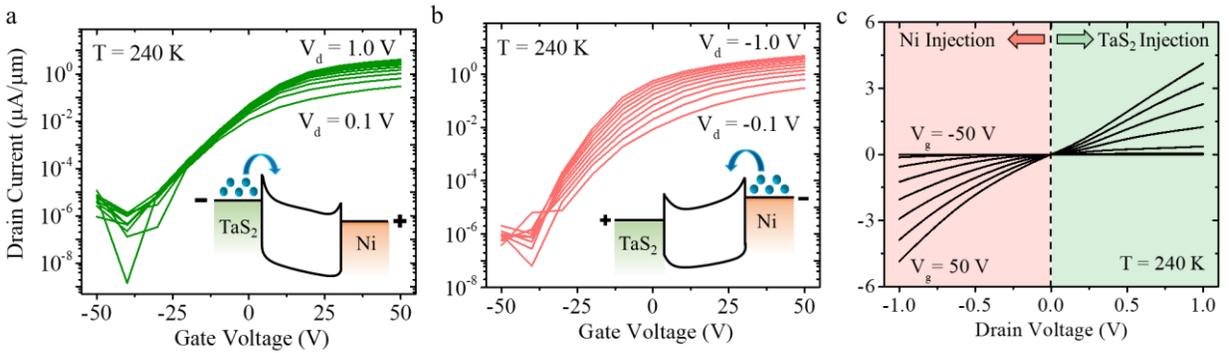

**Figure 3. Current-voltage characteristics of the heterojunction device H23.** Probing terminal 2 and 3 in Fig. 1a-b. (a)-(b) Transfer characteristics at 240 K (T phase) at different drain voltages, with (a) $V_d > 0$ (Ni is under positive bias and TaS$_2$ sourcing electrons) and (b) $V_d < 0$ (TaS$_2$ is under positive bias and Ni sourcing electrons). (c) Output characteristics of the same device.

***Modulating drive current by C-T phase transition:*** We next turn our attention to the control of the carrier injection as the phase of the 1T-TaS$_2$ source undergoes a C-T phase transition. The two different situations are schematically depicted in Fig. 4a-b. Note that, in this heterojunction device, the overall current density through the TaS$_2$ source is smaller compared to the high field case discussed in Fig. 2e (TS12) owing to the series resistance offered by the MoS$_2$ channel, and hence the role of local Joule heating in the phase transition in TaS$_2$ source can be ruled out. This results in a more uniform phase transition in TaS$_2$ source controlled by the external temperature. The measured device current with $V_d > 0$ and $V_d < 0$ are plotted as a function of temperature in Fig. 4c and 4d, respectively. The effect of the TaS$_2$ phase change on the device current manifests as a sharp increase in the drive current for both TaS$_2$ and Ni injection, as indicated by the black arrows. The change in drain current ($I_d$) can be attributed to the change in the series resistance offered by the TaS$_2$ portion of the device due to the C-T phase transition. The fractional enhancement of drive current during phase change is stronger at higher $V_g$ due to reduced MoS$_2$ channel resistance.

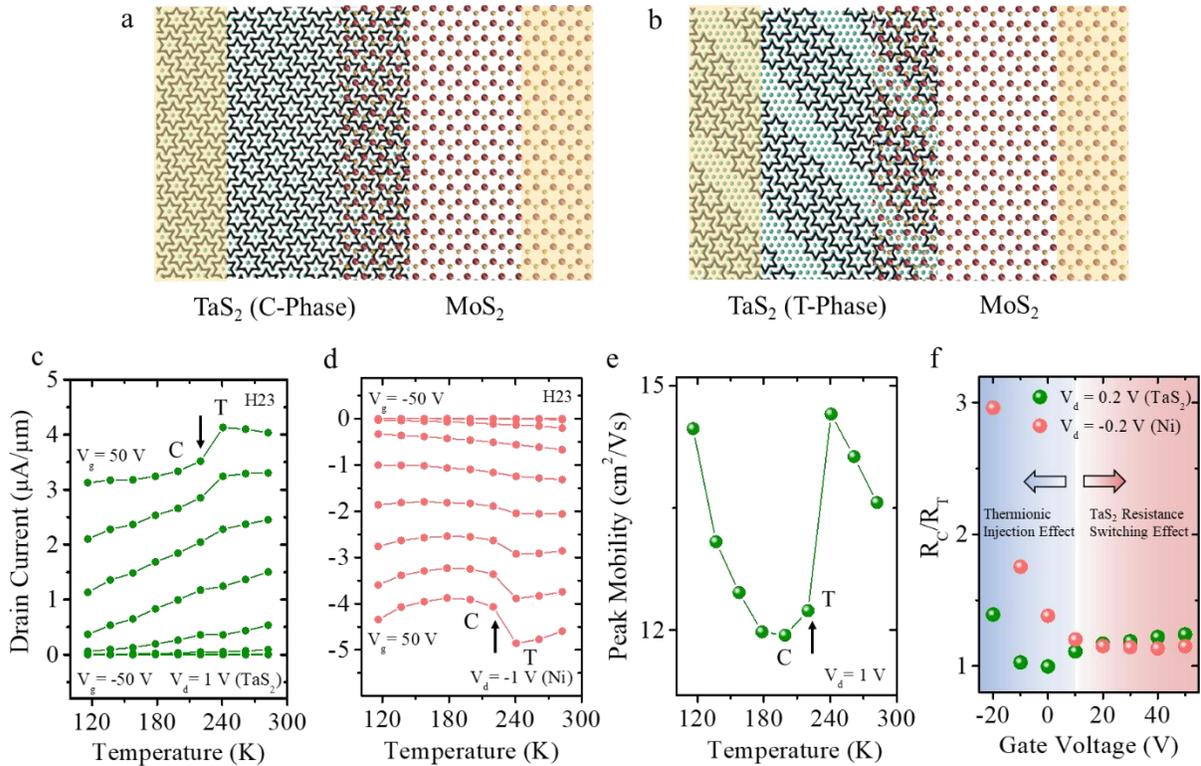

**Figure 4. Drive current impact of C-T phase transition in 1T-TaS$_2$/2H-MoS$_2$ heterojunction device.** (a)-(b) Schematic of the heterojunction with TaS$_2$ in C-phase [in (a)] and in T-phase [in (b)]. (c)-(d) Drive current of device H23 under $V_d > 0$ [in (c)] and $V_d < 0$ [in (d)]. The C-T phase transition temperature of TaS$_2$ is indicated by black arrows. (e) Temperature dependent peak

extrinsic mobility of H23. (f) Resistance ratio ($\rho = R_C/R_T$) dependence on the gate voltage for TaS$_2$ injection ($V_d = 0.2\ V$) and Ni injection ($V_d = -0.2\ V$).

The MoS$_2$ extrinsic electron mobility (i.e., the effect of series resistance in mobility calculation has not been de-embedded) in the heterojunction device has been extracted by using the relation: $\mu = \frac{L}{WC_{ox}V_d} \times \frac{dI_d}{dV_g}$ where $L$ is the channel length, $W$ is the channel width, $C_{ox}$ is the back-gate oxide capacitance. The peak extrinsic mobility decreases with an increase in temperature, as shown in Fig. 4e. Note that the suppression in the TaS$_2$ series resistance manifests itself by the sharp increase in the peak extrinsic mobility at the phase transition temperature, indicated by the black arrow.

Fig. 4f shows the gate voltage dependence of the ratio of the measured resistances in the C-phase and the T-phase ($\rho = R_C/R_T$). Here, we define $R_C$ and $R_T$ as the total resistance measured right before (at 220 K) and right after (at 240 K) C-T phase transition. At large negative $V_g$, the total resistance is governed by thermionic injection over the source-channel barrier and consequently the ratio exponentially increases for the Ni injection case ($V_d = -0.2\ V$) due to an increase in the temperature. However, at large positive $V_g$, where the current injection is dominated by tunneling through the Schottky barrier, the ratio becomes close to unity, but remains larger than 1. In fact, at larger positive $V_g$, the ratio increases with an increase in $V_g$. This effect is slightly more prominent in the TaS$_2$ injection ($V_d = 0.2$) case. Such an increase in the ratio is due to a gradual reduction in the MoS$_2$ channel resistance with $V_g$, and hence the effect due to TaS$_2$ series resistance change becomes more pronounced.

***Phase transition induced SBH modulation and extraction of TaS$_2$ C-phase Mott gap:*** In Fig. 4f, we observe a surprisingly large difference in the ratio $\rho$ in the TaS$_2$ and Ni injection cases when $V_g$ is below threshold voltage. $\rho$ is found to be suppressed and strongly non-monotonic in $V_g$ at large negative $V_g$ for the TaS$_2$ injection case, where the carrier injection is governed by thermionic injection over the Schottky barrier height. This suggests a change in the barrier height at the TaS$_2$/MoS$_2$ interface due to the C-T phase change. To explore the SBH at the TaS$_2$/MoS$_2$ heterojunction, in Fig. 5a, we plot the H23 drain current (in log scale) at $V_d = 0.1$ V as a function of temperature, for different gate voltages. As indicated by the dashed box, the device current is found to be suppressed by the C-T phase transition at low gate voltages, although this effect smears out at higher gate voltage. This suppression of current is in contrary to the current enhancement effect at large $V_g$ in Fig. 4c. Note that at such a low gate voltage, the channel resistance is high, hence TaS$_2$ series resistance or channel mobility do not play any role in such flattening of the device current. This indicates an increase in the SBH at the TaS$_2$/MoS$_2$ contact interface once the phase change in TaS$_2$ sets in. At small $V_g$, the current injection is completely governed by thermionic emission over the TaS$_2$/MoS$_2$ barrier. An increase in the temperature increases the thermionic emission probability, but the C-T phase change abruptly increases the SBH, compensating for the temperature increase effect. Note that, no such current suppression behavior is observed when Ni injects the electrons (i.e., for $V_d < 0$) into MoS$_2$ channel (dashed rectangular box in Fig. 5b). This is due to lack of any phase change of the Ni source, unlike TaS$_2$, hence SBH at the Ni source remains the same before and after $T_{CT}$.

We extract the total effective barrier ($\varphi_B$) offered by the TaS$_2$/MoS$_2$ junction at a given $V_g$ using Richardson equation. As recently proposed[39], we use a modified Richardson equation for such a top contact geometry: $I_d = A^* T^\alpha e^{-q\varphi_B/k_B T}$ where $1 \leq \alpha \leq 1.5$. This differs in the power of $T$ from the Richardson equation typically used for the interface between a metal and a conventional bulk semiconductor owing to a change in the dimensionality. $\varphi_B$ is extracted from the slope of $\log(\frac{I_d}{T^\alpha})$ versus $\frac{q}{k_B T}$ as depicted in Fig. 5c. One could clearly observe the abrupt increase in the slope as the phase transition happens, indicating an increase in $\varphi_B$ in the T phase. The extracted $\varphi_B$ is plotted in Fig. 5d as a function of $V_g$ for both C and T phases. We note here that, the extracted $\varphi_B$ is not very reliable at large positive $V_g$ due to strong tunneling current and mobility degradation with temperature, which tend to underestimate the extracted effective barrier height.

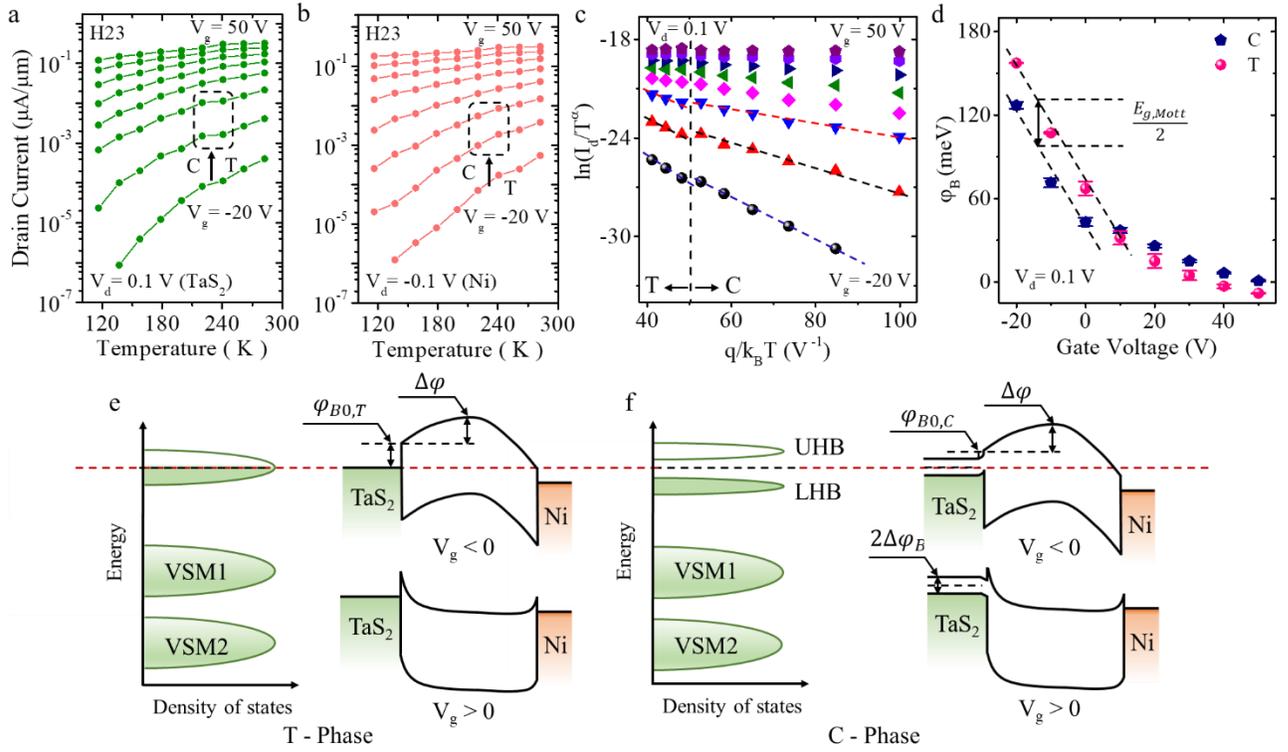

**Figure 5. Phase dependent Schottky barrier height at 1T-TaS$_2$/2H-MoS$_2$ interface of H23 and Mott gap estimation in C-Phase of 1T-TaS$_2$.** (a) Drain current as a function of temperature under small positive drain bias (TaS$_2$ injection). The dashed box indicates the suppression of drain current by the C-T phase transition under low $V_g$, which smears out at higher $V_g$. (b) Drain current as a function of temperature under small negative drain bias (Ni injection), showing no such suppression in the dashed box. (c) Richardson plot (with $\alpha = 1$) at different gate voltages in two different phases, indicating a C-T phase transition driven change in slope at negative $V_g$ due to change in barrier height. (d) Extracted barrier height plotted as a function of gate voltage in two different phases. The dashed parallel lines indicate linear increase of barrier height for larger negative $V_g$. (e)-(f) Schematic representation of the barrier height increase mechanism from C-phase to T-phase.

In Fig. 5d, extracted barrier height is found to increase linearly with decrease in $V_g$ when $V_g$ is small. The *'knee point'* in the curve, where the barrier height deviates from linearity, is indicative of the true Schottky barrier height $\varphi_{B0}$ (i.e. under flat band condition) of the TaS$_2$/MoS$_2$ interface[51]. This is much lower compared with the difference between the work function of 1T-TaS$_2$ (5.2 eV)[52] and the electron affinity of multi-layer MoS$_2$ (4 eV)[53,54]. This suggests a strong Fermi level pinning at the interface, close to the conduction band edge of MoS$_2$ – supporting the excellent carrier injection efficiency through the interface. Such a strong Fermi level pinning is a unique feature of TaS$_2$/MoS$_2$ vdW contact, suggesting the vdW gap does not efficiently suppress the evanescent wave function of the TaS$_2$ states, likely resulting in metal induced gap states (MIGS)[55] in MoS$_2$.

Note that the increase in barrier height with negative $V_g$ in C and T phases can be fitted by two parallel lines, as shown in Fig. 5d. This observation indicates that the band bending ($\Delta\varphi$) in MoS$_2$ is similar in both cases at large negative $V_g$. This is schematically explained in Fig. 5e-f, which allows us to write the total barrier as: $\varphi_{B,p}(V_g) = \varphi_{B0,p} + \Delta\varphi(V_g)$, where $\varphi_{B0,p}$ is the true SBH of the TaS$_2$/MoS$_2$ interface (under flat-band condition) and $p \in \{C, T\}$ represents the phase of TaS$_2$. Assuming a symmetric Mott gap opening, $\Delta\varphi_B$ provides an estimate of the Mott gap in TaS$_2$ in the C phase:

$$\Delta\varphi_B = \varphi_{B,T} - \varphi_{B,C} \approx \varphi_{B0,T} - \varphi_{B0,C} \approx \frac{E_{g,Mott}}{2}$$

The vertical separation of the dashed fitting lines in Fig. 5d is an indicator of $\Delta\varphi_B$. The extracted Mott gap of TaS$_2$ in the C-phase is estimated to be $E_{g,Mott} \approx 71 \pm 7$ meV. This is in reasonable agreement with reported numbers in literature from different optical techniques viz. infrared reflectivity[20,21], time resolved photoemission spectroscopy[9,22], angle resolved photoemission spectroscopy[10,23,24] and angle resolved inverse photoemission spectroscopy[25] – providing an independent verification using pure electrical transport method.

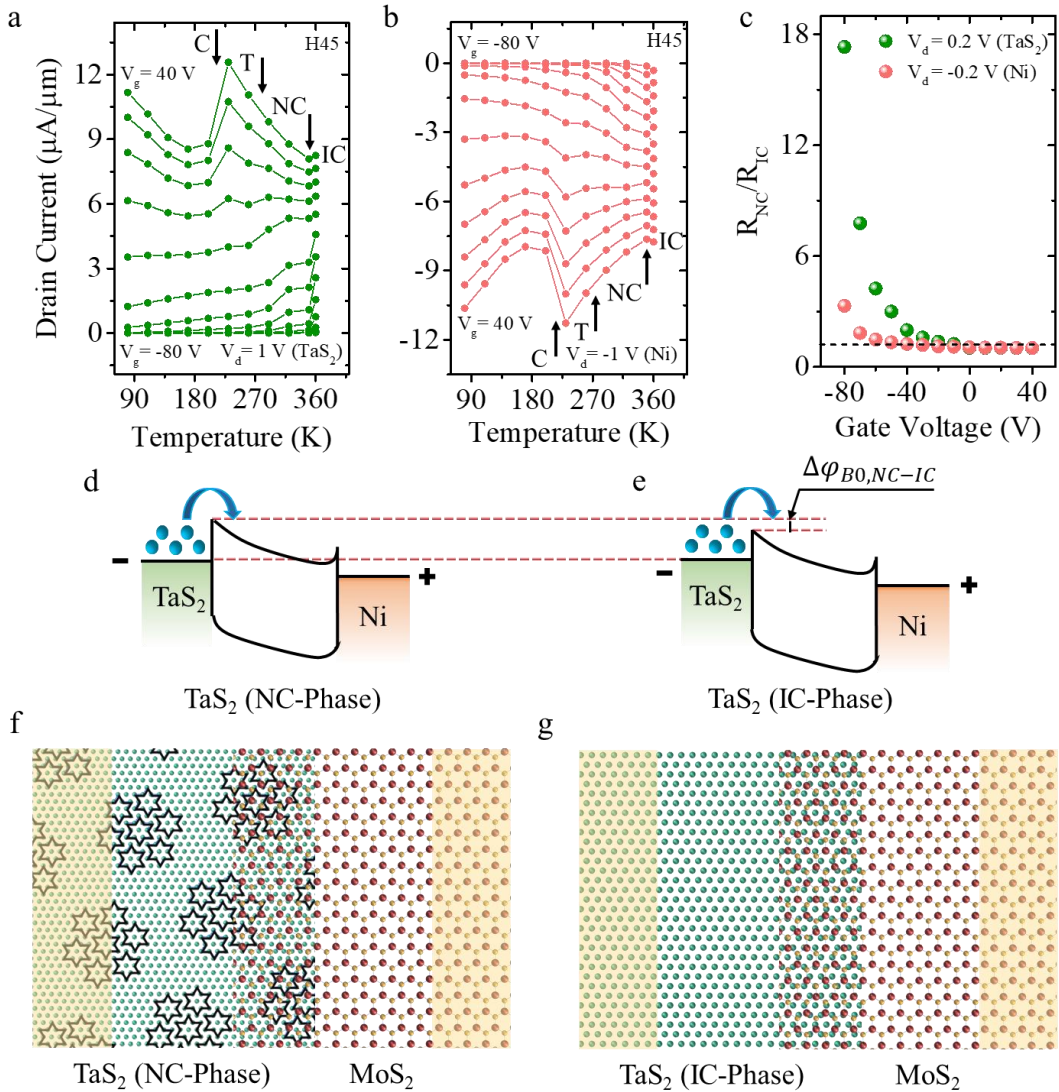

**Figure 6. Enhanced resistance switching during NC-IC phase transition in 1T-TaS₂/2H-MoS₂ heterojunction device H45.** (a)-(b) Drive current of device H45 under $V_d > 0$ [in (a)] and $V_d < 0$ [in (b)]. The different phase transition temperatures of TaS₂ are indicated by black arrows. (c) Resistance ratio ($R_{NC}/R_{IC}$) as the function of gate voltage for TaS₂ injection (in greed symbols, $V_d = 0.2\ V$) and Ni injection (in orange symbols, $V_d = -0.2\ V$). The dashed line indicates 1T-TaS₂ control. (d)-(e) Schematic for SBH height reduction from NC [in (d)] to IC-phase [in (e)]. (f)-(g) Schematic of the heterojunction with TaS₂ in NC-phase [in (f)] and in IC-phase [in (g)].

*Enhancing resistance switching during NC-IC phase transition:* Fig. 6a-b depict the temperature dependent drain current characteristics from another device (H45) possessing a lower threshold voltage than H23, and the device is driven deep into the inversion by increasing the overdrive voltage. An optical image of the device H45, along with the control TaS₂ transport characteristics are provided in **Supplemental Material S2**[40]. In this device, we could modulate the drive current by as much as ~40% through the C-T phase change under large gate overdrive condition. The strong suppression of the drive current at higher temperatures (beyond 223 K) is due to the temperature induced mobility degradation effect. The temperature dependent peak extrinsic mobility extracted from H45 is shown in **Supplemental Material S2**[40]. When we drive the temperature of the device up to 360 K, which is beyond the NC-IC phase transition temperature at 353 K, we observe that the NC-IC phase transition manifests as a step jump in the drive current, both for TaS₂ and Ni injection cases as in Fig. 6a and b, respectively. The corresponding resistance switching ratios ($\frac{R_{NC}(T=350\ K)}{R_{IC}(T=360\ K)}$) at $V_d = \pm 0.2$ V are plotted in Fig. 6c. For reference, we also show the ratio for the TaS₂ control as a dashed line in the same plot. For TaS₂ injection case ($V_d = 0.2$), the switching ratio is a strong function of $V_g$ and is remarkably large at negative $V_g$ reaching a value of 17.3 at $V_g = -80$ V, which is 14.5 times higher than the TaS₂ control device. On the other hand, the ratio remains a weak function of $V_g$ for Ni injection case and remains close to the value of TaS$_2$ control.

These observations point to a suppression of the SBH at the TaS$_2$/MoS$_2$ interface due to the NC-IC phase transition, as schematically depicted in Fig. 6d-g. Consequently, for electron injection from TaS$_2$ source, the current modulation is much higher, while for Ni injection, we only get small effect due to a change in the series resistance of TaS$_2$ during the phase transition. While the origin of such a change in SBH requires further investigation, it is likely that during the NC-IC phase transition, as the hexagonal David-star clusters are broken (Fig. 6f-g) to increase conductivity, there is a more pronounced effect of the MIGS from TaS$_2$ into the bandgap of MoS$_2$. This causes the Fermi level to be pinned closer to the conduction band edge of MoS$_2$, reducing the SBH.

**Conclusion:**

In conclusion, we have demonstrated a low-barrier efficient electrical contact between 1T-TaS$_2$ source and 2H-MoS$_2$ channel, which is promising for "all-2D" flexible electronics. Along with the usual conductivity switching of 1T-TaS$_2$ during different phase transitions, we discovered that these transitions also bring about a change in the Schottky barrier height at the 1T-TaS$_2$/2H-MoS$_2$ interface. The phase transition driven resistance switching ratio of the heterojunction thus shows a large modulation that can be controlled by an external gate voltage. This enhancement and additional gate control provide an unprecedented opportunity for boosting different device applications which exploit such phase transition induced resistance switching, such as broadband photodetection, neuromorphic circuits, negative differential conductance and fast oscillator.

**Acknowledgements:**

K. M. acknowledges the support a grant from Indian Space Research Organization (ISRO), grants under Ramanujan Fellowship, Early Career Award, and Nano Mission from the Department of Science and Technology (DST), Government of India, and support from MHRD, MeitY and DST Nano Mission through NNetRA.

**Conflicts of Interest:**

The authors declare no conflict of interest.

465–468 (1984).

**Supplemental Material:**

**Gate controlled large resistance switching driven by charge density wave in 1T-TaS$_2$/2H-MoS$_2$ heterojunction**

Mehak Mahajan[∥], Krishna Murali[∥], Nikhil Kawatra, and Kausik Majumdar[*]

Department of Electrical Communication Engineering, Indian Institute of Science, Bangalore 560012, India

[∥]These authors contributed equally

[*]Corresponding author, *email: kausikm@iisc.ac.in*

**Supplemental Material S1:**

**Output characteristics of MoS$_2$ channel with Ni contact on both sides**

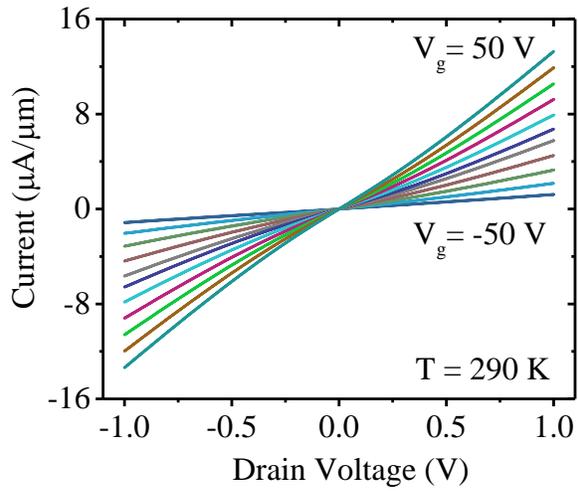

**Figure S1.** I$_d$-V$_d$ characteristics of MoS$_2$ channel with Ni contact on both sides for different gate voltages, at T = 290 K, with channel length of 2.8 μm.

**Supplemental Material S2:**

**Characteristics of TaS$_2$ control and heterojunction from device H45**

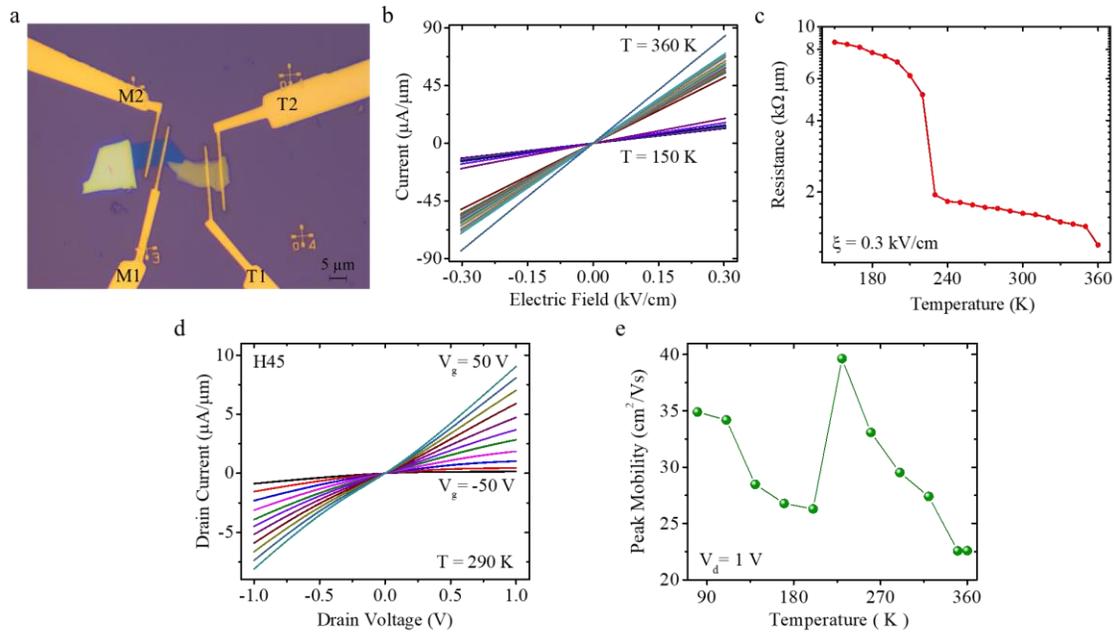

**Figure S2.** (a) Optical image of the heterojunction device H45. T1 and T2 are the contacts to the TaS$_2$ flake, while M1 and M2 are contacts to MoS$_2$. Scale bar is 5 µm. (b) Current versus electric field characteristics of TaS$_2$ control (probed between T1 and T2) at different temperatures and low electric field condition. (c) The resistance versus temperature plot extracted from (b) clearly indicating the C-T transition around 220 K and the NC-IC phase transition around 350 K. (d) Output characteristics of the heterojunction H45 (probed between T1 and M1) at 290 K. (e) Temperature dependent peak extrinsic mobility, as extracted from H45.